\documentclass[aps,pre,onecolumn,superscriptaddress,floatfix]{revtex4}
\usepackage{epsf}
\usepackage{amsmath}
\usepackage{amssymb}
\usepackage{color}
\usepackage{graphicx}
\usepackage{subfigure}

\begin{document}

\title{ Effects of the Substitutions and Vacany  Defects on the Stable monolayer structures of Black and Blue  Arsenic Phosphorus: A First Principles Study}

\author{Zhaleh Benam}
\email[E-mail: ]{zhaleh.benam@gmail.com}
\affiliation{Department of Physics Engineering, Faculty of Engineering, Ankara University,
Tandogan, 06100 Ankara, Turkey}

\author{Handan Arkin}
\email[E-mail: ]{holgar@eng.ankara.edu.tr}
\affiliation{Department of Physics Engineering, Faculty of Engineering, Ankara University,
Tandogan, 06100 Ankara, Turkey}

\author{Ethem  Akt\"{u}rk}
\email[E-mail: ]{ethem.akturk@adu.edu.tr}

\affiliation{Department of Physics, Adnan Menderes University, 09100 Ayd{\i}n, Turkey}
\affiliation{Nanotechnology Application and Research Center, Adnan Menderes University, 09100 Ayd{\i}n, Turkey}

\begin{abstract}
We investigate two dimensional monolayer  structures of arsenic phosphorus ( AsP) by means of first principles plane wave method within
  density functional theory using the generalized gradient approximation. 
Two structures of arsenic phosphorus are taken into account which are called
buckled (B-AsP) and puckered (Pu-AsP). 
From our numerical calculations we predicted  the optimized states of 
 a single-layer buckled and puckered 
 honeycomb like allotropes of AsP. 
We demonstrate that buckled  and puckered arsenic phosphorus are  
  semiconducting  single layers with band gap energies 1.858 eV and  0.924 eV,  respectively. 
Especially,  we have found that the puckered monolayer has a direct band gap. 
Furhermore, we examined the  electronic and magnetic properties of various
 kind of point defects including single and double
vacancies, antisite and substitutions on both buckled and puckered 
arsenic phosphorus. We conclude that AsP monolayers can attain 
magnetic state with vacancies and the semiconductor nature can turn into 
metal with diverse defect types. In considered 
system also the band gap of B-AsP can be tuned from metal character 
to 1.858 eV and for Pu-AsP from metal to 0.924 eV which 
these ranges are especially important for radar and 
atmosferic applications.   

\end{abstract}


\maketitle

\section{Introduction}

The two dimensional monolayer honeycomb structures have exceptional 
properties and thus attracted a growing interest with theirs promising 
applications in nanotechnology. 
Recent studies in nanoscale physics have aimed the discovering new monolayer materials and 
revealing their properties under different conditions. Among these materials, graphene, 
which is a single layer honeycomb structure of carbon, has attracted great interest in various 
fields ranging from electronics to biotechnology due to its unique properties, such as 
high mechanical strenght, chemical stability, and exceptional ballistic conductance \cite{Geim, Novoselov}.
The search for the contender of graphene has led to the prediction/synthesis of new single-layer, 
crystalline nanostructures, which do not exit in nature. Silicene, germanene, stanene, 
namely graphene analogs of group IV or V elements \cite{Durgun, Cahangirov, Vogt, Ozcelik, Xu}; h-BN analogs of group IV-IV, III-V,
 and II-VI compounds \cite{Sahin}; graphyne \cite{Tongay, Malko, Ozcelik2} and the auxetic piezoelectric 2D material 
silicatene with negative Poissons ratio \cite{Ozcelik3, Xu2}  have been actively studied. Additionally, transition-metal dichalcogenides
 have proven to be superior than graphene in specific optoelectronic applications \cite{Joensen, Coleman, Ataca1, Ataca2}. 
Especially, the  silicene and
germanene \cite{Cahangirov} have attained
significant interest due to their nonmagnetic semimetallic behavior
 with linearly crossing $\pi$ and $\pi^*$ bands at Fermi
level which is similar to graphene and due to their compatibility with silicon-based micro-electronic industry.
In addition, two dimensional silicon carbide (SiC) and germanium carbide
(GeC) honeycomb structures have been reported theoretically
as stable \cite{Bekaroglu,Sahin,Sun,Xu3}.
However, the electronic structure of graphene, and silicene, germanene
as well have zero band gap, which restrain the ability and applications
 of them.

More recently, the fabrication of field effect transistors using micrometer-sized flakes 
consisting of two and three layers of black phosphorus \cite{Li}  and their theoretical analysis \cite{Zhu,Low}  
revealing the stability of its single-layer allotropes brought group V elements into the focus. 
In contrast to the zero band gap of honeycomb lattices
of group-IV elements, monolayer
honeycomb structures of group-V elements such as Phosphorene \cite{PRL},
Arsenene \cite{PRBarsenene}  and Antimonene \cite{PRBakturk} are
direct and indirect band gap semiconductors,
 respectively and they  were  found as
stable \cite{PRL, PRBarsenene, PRBakturk} and the mechanical  and
electronic \cite{AngeChemie}  
properties  were investigated.
Because  of direct or indirect band gap
 properties, these honeycomb monolayer structures  can be promising
materials for digital circuits and light-emitting diodes.

To date, these 2D layered materials have covered a wide electromagnetic spectral range including metals, 
semimetals, semiconductors, and insulators. Specifically, monolayer graphene is a semimetal which 
exhibits a zero band gap in its natural states and strong electrical field can be used in order 
to create a  band gap in bilayer graphene \cite{Zhang}.
Few-layer and/or monolayer structures of group V elements, including phosphorene and arsenene, 
are emerging as promising candidates for two-dimensional (2D) electronic materials application \cite{ Narita, Maruyama, Liu, Zhu2,  Liu2}. 
Different from semimetallic graphene, these systems display a nonzero band gap while still 
maintaining a relatively high carrier mobility \cite{Zhang2, Koenig, Gomez, Xia}. 
One of the newest members of the 2D layered material family is 
black phosphorus, which is a semiconductor exhibiting a moderate band gap of 
around 0.3 eV in its bulk form and up to 2.2 eV as in a monolayer \cite{Wang}, 
further pushing the band gap of single layered materials 
to the middle-wavelength infrared regime. Therefore, a broad spectral 
range has now been covered by 
the above mentioned layered materials. Although a wide range in 
electromagnetic spectrum has been 
covered by these 2D monolayer materials, there is still a big need to different  
smart materials because of  the light radars and atmosferic applications 
in the longwavelength spectrum range.
On the other hand, because of its direct band gap, black phosphorus is 
currently considered the 
best material on-chip dedectors. Besides all of these advantages, few-layer black phosphorus  
samples demonstrate  high reactivity, resulting in a fast degradation of crystals under ambient 
conditions and while being chemically much less reactive, layered gray arsenic also display 
a significant band gap. In order that reason, the combination of black 
phosphorus and gray arsenic 
may exhibit  phase coexistence that should bring an unexpected richness in both structural and 
electronical properties, in comparison to pristine of phosphorene and arsenene, has already 
been shown by the end of year 2015 \cite{Tomanek}. This structure has
 been called black arsenic phosphorus and 
the honeycomb structure which can also be exist in nature is 
called blue arsenic phosphorus.

Although both buckled and puckered arsenene monolayers are stable 
structures which are shown
by phonon spectrums and with theirs cohesive energies, they possess 
indirect band gaps of 
0.831 eV and 1.635 eV, respectively \cite{PRBarsenene}. 
In this paper, motivating by recent studies of phosphorene and arsenene, 
our aim is modifying
the indirect gap of arsenene by making black arsenic phosphorus 
or in other words 
by substitution of arsenic atoms to phosphorene. 
Additionally,  we are thinking that  the electronic properties of these new phases of 
arsenic phosphorus can be tuned further and can be  functionalized by 
 substitution and vacany processes.
Because previous experimantal and theoretical studies have 
proven that bare 2D monolayer structures
can be functionalized through point defects  to attain crucial physical and chemical properties
for diverse applications.

On this object, we present ab initio calculations of the arsenic phosphorus
 allotropes in the frame of first principles calculations. 
Beside Si, Ge, Sn monolayers and with the development of experimental technology,
AsP monolayer  have the advantage of manufacturing by exfoliating crystalline
black phosphorus which has been shown recently by Liu et. al. \cite{Adv.Mat.2015}.
 In this study, a new synthetic route was used for the synthesis of AsP layers which uses 
gray arsenic and red phosphorus  as starting materials. 
Therefore AsP is chosen as the research objective of interest here.
We investigate the stability and electronic properties of the 2D honeycomb 
structures. We have found that two honeycomb structures called buckled (B-AsP) and puckered (Pu-AsP) arsenic phoshporus  are stable and the 
puckered AsP structure has a direct band gap. 
Furthermore we  focus on  investigating the effects of  the point 
defects in the  physical properties
of black and blue arsenic phosphorus monolayers by 
first principles calculations.
 In all structures the geometric optimization are done and the recontruction,
the effects on the electronic properties have been  discussed.
In summary, the works discussed  in this letter  may  open a new page in the field of material
science by predicting new “smart surfaces”.

The rest of the paper is organized as follows. In Section
II the calculation methods  are  described in detail. Then, Section III
contains the results and discussion. Finally, Section
IV concludes the paper with a summary of our findings.

\section{Calculation Methods}
The structural stability and the electronic properties of the Arsenic 
Phosphorus  monolayer structures  have been investigated 
by means of first-principles  plane-wave calculations
within density functional theory (DFT) using projector augmented wave (PAW) potentials \cite{paw}.
The exchange-correlation potential is approximated by generalized 
gradient approximation  
(GGA) using Perdew-Burke-Ernzerhof (PBE) parametrization \cite{pbe}. 
All numerical results have been obtained
by using Quantum Espresso software \cite{pwscf}. 
A plane-wave basis set with the kinetic energy cutoff
$\hbar^2(\textbf{k+G})^2/2m=$ 65 Ry is used. Pseudopotentials with 
 $4s^{2}3d^{10}4p^{3}$ and $3s^{2}3p^{3}$ valence electron configurations 
for As and P atoms were used, respectively.
 All structures are treated using periodic boundary conditions.

The Brillouin zone has been sampled by (21$\times$21$\times$1) and 
(7$\times$7$\times$1) special mesh points in
\textbf{k}-space by using  Monkhorst-Pack scheme \cite{monk} 
for (1$\times$1) and (4$\times$4) AsP cells, respectively. 
All atomic positions and lattice constants are optimized by 
using BFGS quasi-Newton algorithm \cite{BFGS} where the total energy 
and forces are minimized. The convergence criteria for energy is 
chosen as $10^{-8}$ Ry between two consecutive steps. The maximum 
Hellmann-Feynman forces acting on each atom is less than 0.001 Ry/au upon 
ionic relaxation. 
The maximum pressure on the unit cell is less than 0.5 kbar. 
Gaussian type Fermi-level smearing method  is used with a smearing width
of 0.1 eV.

\section{Results and discussion}

\subsection{Structural properties of the monolayers}
\begin{figure*}[t]
\hspace*{-3.95cm}\includegraphics[scale=0.22]{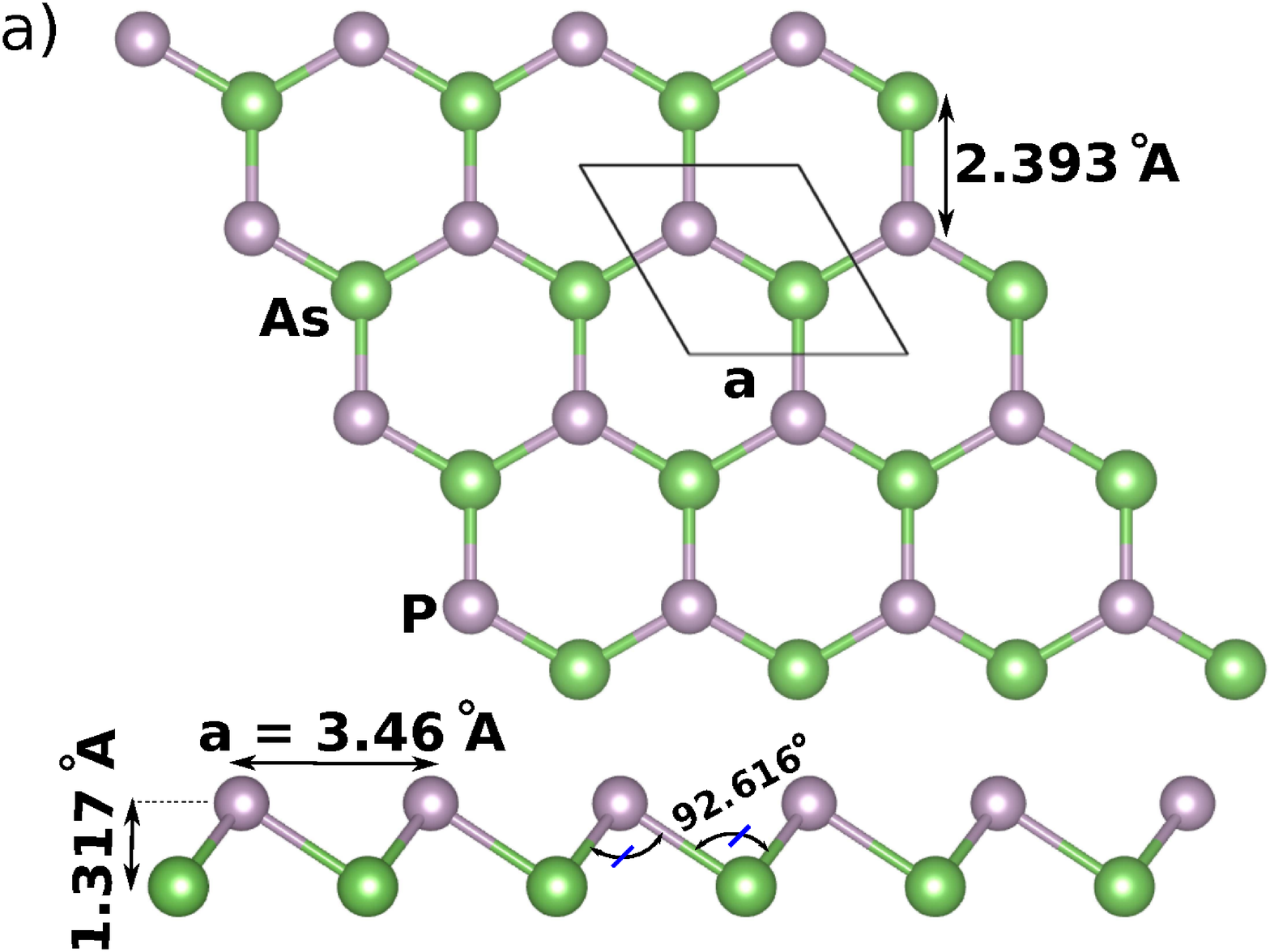} 
\includegraphics[scale=0.27]{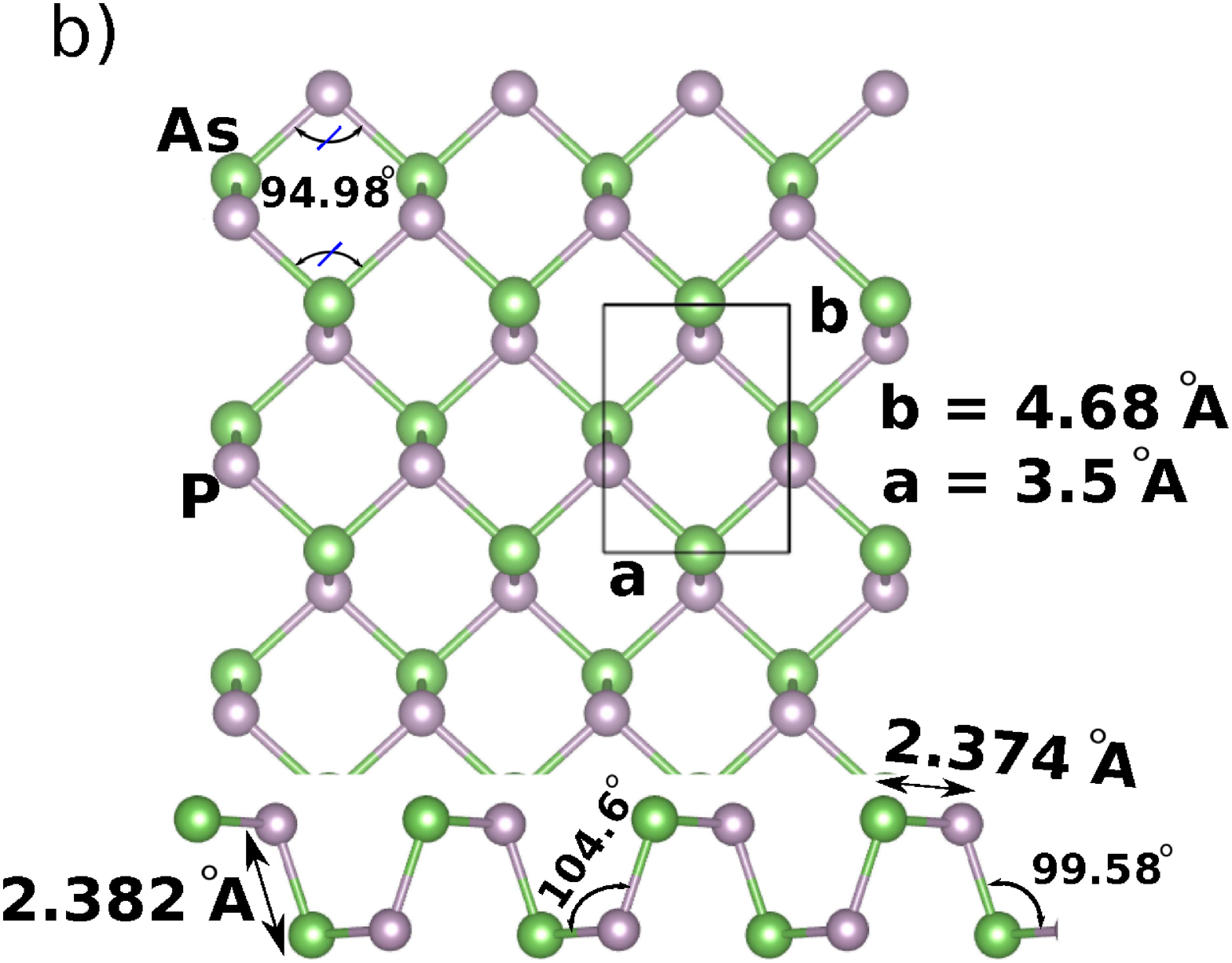} 
\caption{Side view and top view of the structures of a) buckled B-AsP and b) puckered Pu-AsP 
   }
\label{fig:1}
\end{figure*}

Graphene is a planar honeycomb structure since it form pure $sp^2$ hybridization. On the
other hand except graphene the other elemental  honeycomb monolayers form 
buckled structures for example the silicene and germanene are the most popular ones.
In addition, which has recently predicted, the phosphorene made of black phosphorus has
a puckered structure and its buckled structure namely blue phosphorus has been 
investigated by first principles study \cite{PRL}.
From this purpose, we take two  structures of arsenic phosphorus into account and 
 show the optimized 
geometric structures for the cases a) buckled b) puckered  in Fig.\ref{fig:1}
 from side view and from top view. 
The structural parameters corresponding to the two different structures
 which are under consideration in this study are given  in Table\ref{tab:1}. 
The lattice constants of optimized orthorhombic structure for Pu-AsP
in equilibrium are $a=3.50 $ and $b=4.68$ \AA\, which are about 4 percent   longer 
than the lattice vectors of black phosphorene
 and shorter than the lattice vectors of black arsenene. 
For the  hexagonal structure 
  B-AsP  in equilibrium are $a=b=3.46$ \AA\ . 
 The B-AsP is again 4 percent   longer than the blue phosphorene and 
shorter than blue arsenene. The bond distance between As and P atoms are
$d_{1As-P}= 2.374 $ \AA\ and $d_{2As-P}= 2.382 $ \AA\ for Pu-AsP, $d_{As-P}= d=2.393$ \AA\
 for B-AsP. We found that the bond lengths of both
puckered and buckled monolayers are smaller than its counterpart of  arsenenes and slightly
bigger than  black- and blue-phosphorene.

 To prove the stability of the monolayers, we calculated theirs cohesive 
energies defined as 
$E_{coh} = ( nE_{As} + nE_{P} - nE_{AsP} ) / 2n$, where $E_{As}, E_{P}$ and 
$E_{AsP}$  are the total energies of single As atom, a single P atom, and a As-P pair 
in the AsP monolayer, respectively. 
 The  puckered and buckled monolayer structures have a cohesive energies  −4,782 eV and −4,802 eV respectively per As-P pair  
 and these values are  very similar for the different monolayer structures which are
stable and manufactured successfully. These values are clear evidence that AsP 
monolayers  are strongly bonded networks.
To check the structure optimizations, we increase the degrees of freedoms to 4x4 cells
which we got the same geometry and cohesive energy.

In mean, we investigate here a previously unknown structure of arsenic  
phosphorus  which called buckled structure. 
This structure is a honeycomb structure like graphene and it is as 
stable as puckered structure. 
Whereas these structures is hexagonal structures closely related to graphene,
 the main advantage of these structures are those  band gaps which  are approximately 
in the range  
of  1.0 - 2.0 eV. 
Although the band  gaps  are typically underestimated with the DFT calculations from the 
literature the general trend is confirmed,
and the black- and blue arsenic phosphorus have  larger band gaps than the 2D counterparts 
where aditionally the  
puckered structure has the advantage of direct band gap. 
Because  the band gaps of most explored 2D materials exhibite 
moderate band gaps around  0.3 eV, 
which greatly restricts applications based on semiconductor 
and optoelectronic device technology. 

\subsection{Electronic structure}
\begin{figure*}[t]
\includegraphics[scale=0.60]{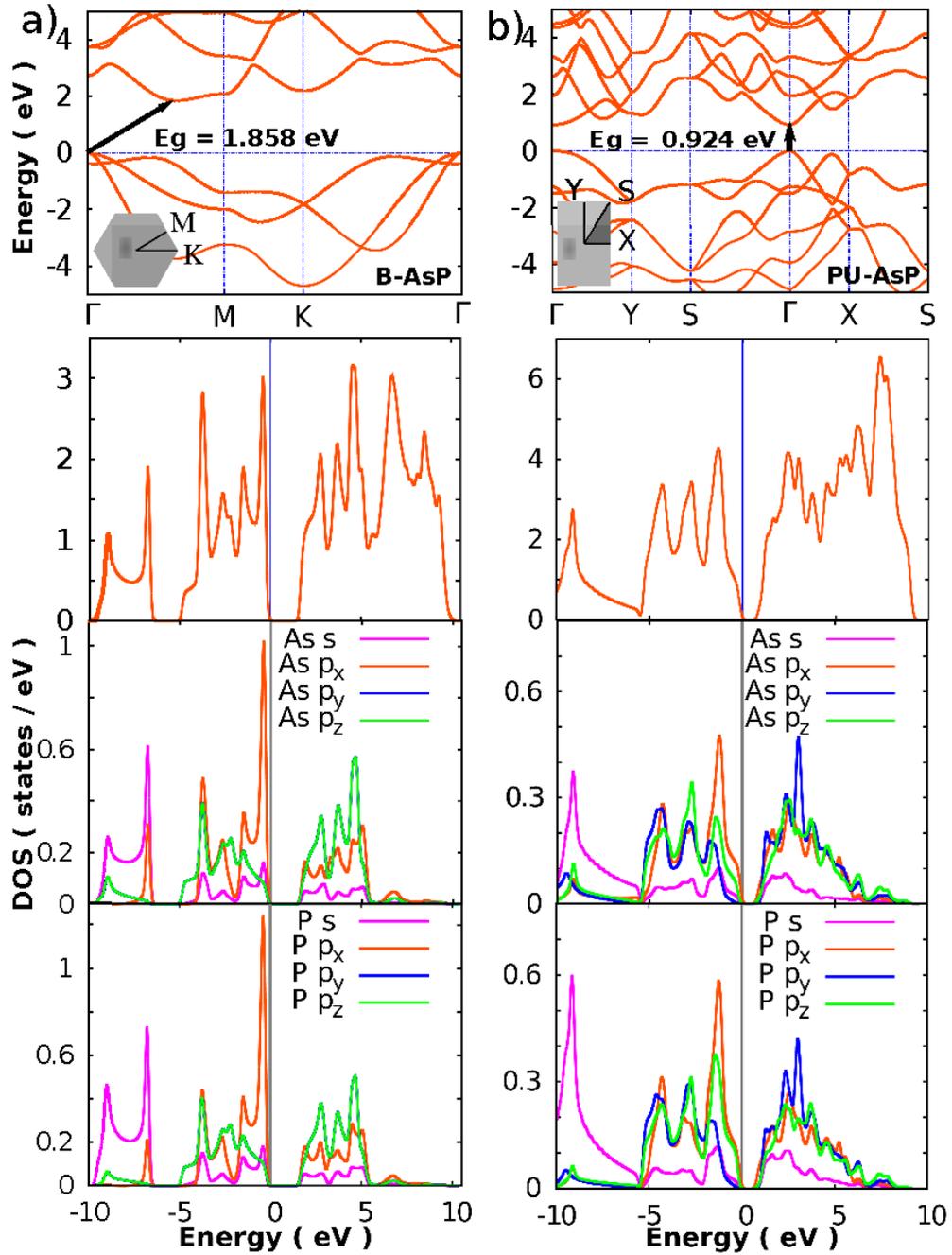}
\caption{Electronic band structures, corresponding total (DOS) and partial (PDOS) density 
of states for the monolayer structures  of 1x1 a)  buckled B-AsP and b) puckered Pu-AsP. 
Zero of energy is set to Fermi level shown by dashed line. 
(For interpretation of the references to color in this figure legend, the reader is referred to the web version of this article.) }
\label{fig:2}
\end{figure*}

Here we carried out the electronic band structure and 
in order to understand the 
contributions of atomic orbitals to the elctronic states we have plotted the 
corresponding total (DOS)  and partial (PDOS) density of states 
for the  buckled and puckered  monolayers  as shown in Fig.\ref{fig:2}a, Fig.\ref{fig:2}b, 
respectively. 
From the electronic band structures it is obviously seen that puckered AsP is direct band 
gap semiconductor with a band gap of  0.924 eV whereas buckled AsP is indirect band gap 
semiconductor with a band gap  of  1.858  eV. It is observed  from the PDOS plots that the 
electronic states  both in puckered and buckled structures have contributions 
from $s$ and $p$ orbitals near the Fermi energy, but the $s$ orbital 
contribution is very small compared to 
$p$ orbital contribution. This is very well known feature of monolayer 
honeycomb systems such as germanene, silicene and phosphorene,
where the $sp^2$-like bonds form a nonplanar honeycomb structures different than graphene.  
An indirect band gap in buckled AsP resembles that of buckled (blue) phosphorene and arsenene. 
The valence band maximum lies at the $\Gamma$-point, and the conduction band minimum lies along 
$\Gamma-M$ direction. Same situation are also seen both in buckled phosphorene and arsenene. Only
the band gap is higher than arsenene in our case. But we could not exactly compare our value
with the indirect band gap of buckled phosphorene, because its band gap was  given as approximate
value \cite{PRL}.   
On the other hand, the direct gap semiconducting behaviour of puckered AsP is distincly different
from the indirect band gap behaviour of puckered arsenene. Its looks like mostly to puckered 
phosphorene. Because both structures have direct band gaps at the $\Gamma$-point 
with band gaps of
 0.91 eV for puckered phosphorene and 0.92 eV for puckered AsP, respectively.
But, additionally we note that,  puckered AsP structure have two seperate valence band and conduction
band edges near the Fermi level. The  maximum of the valence band and the minimum of the conduction
band occur along $\Gamma-Y$ direction and at the $\Gamma$-point which both are direct edges.
This fact is distincly different also from the electronic band structure of 
phosphorene which have 
only direct edge at the $\Gamma$-point.

\subsection{Investigation of the Defects structures}
To investigate the effects of the defects we optimized 4x4 supercells of the buckled and puckered AsP monolayers.
Such size for the supercells provide sufficient distance for the examination of the defects. As presented in
Fig.\ref{fig:1} the lattice constant 
and the bond length of the  4x4 supercell 
  buckled AsP  which
contains 16 As and 16 P atoms, are found to be $ a= b = 13.84 $ \AA\  and $d_{As-P} = 2.393 $ \AA\  respectively. The cohesive
energy is found to be the same with the unit cell. The puckered AsP 4x4 super cell contains 32 As atoms and 32 P atoms.
The lattice constants for this structure are  $a=14.00 $ \AA\ and $b=18.72$ \AA\ and the bond length is $d_{As-P} = 2.382$ \AA\
same with the unit cell.
As illustrated in Fig.\ref{fig:2} valance band maximum and conduction band minimum consist main contribution   of p orbitals of As and P atoms near Fermi level.
Single atom vacancies and antisite  defects are dominate defects 
during the fabrication of single layer materials, 
or some experiment can cause these defects. The presence of these defects can change 
electronic and magnetic properties of 
the materials. We examine in this section that introducing various defects into 
buckled and puckered AsP monolayer
can cause to crucial changes in the electronic and magnetic properties.  
Firstly, all defected structures are subjected to geometrical optimization and we calculated their cohesive energies which
the expression is given before and 
formation energy for the substitution defect  $X_Y$, where the atom Y occupies the site of atom X, 
by using the equation below,

$E_{form} =  E_{substrate + X_Y} - E_{substrate} + E_{Y}  - E_{X} $. \\

Here, the energy for the single As and P atoms are calculated from the As-rich structure 
and from the P-rich structure, respectively. We take arsenene 
monolayer as As-rich structure and phosphorene monolayer for the P-rich structure. 
The considered various defects and the label of them are given as follows:

\begin{itemize}
\item Vacancy: An atom As (P) subtract from its position in the monolayer network labelled as $V_{As} (V_P)$. 
\item Substitution: When As(P) atom placed instead of P(As) atom labelled as $As_P (P_{As})$.
\item Antisite: When the position of neighboring As and P atoms are exchanged ($As\leftrightarrow P$). 
\end{itemize}

\subsection{Defect Structures on B-AsP monolayer}
\begin{figure*}[t]
\hspace*{-2.50cm}\includegraphics[scale=0.85]{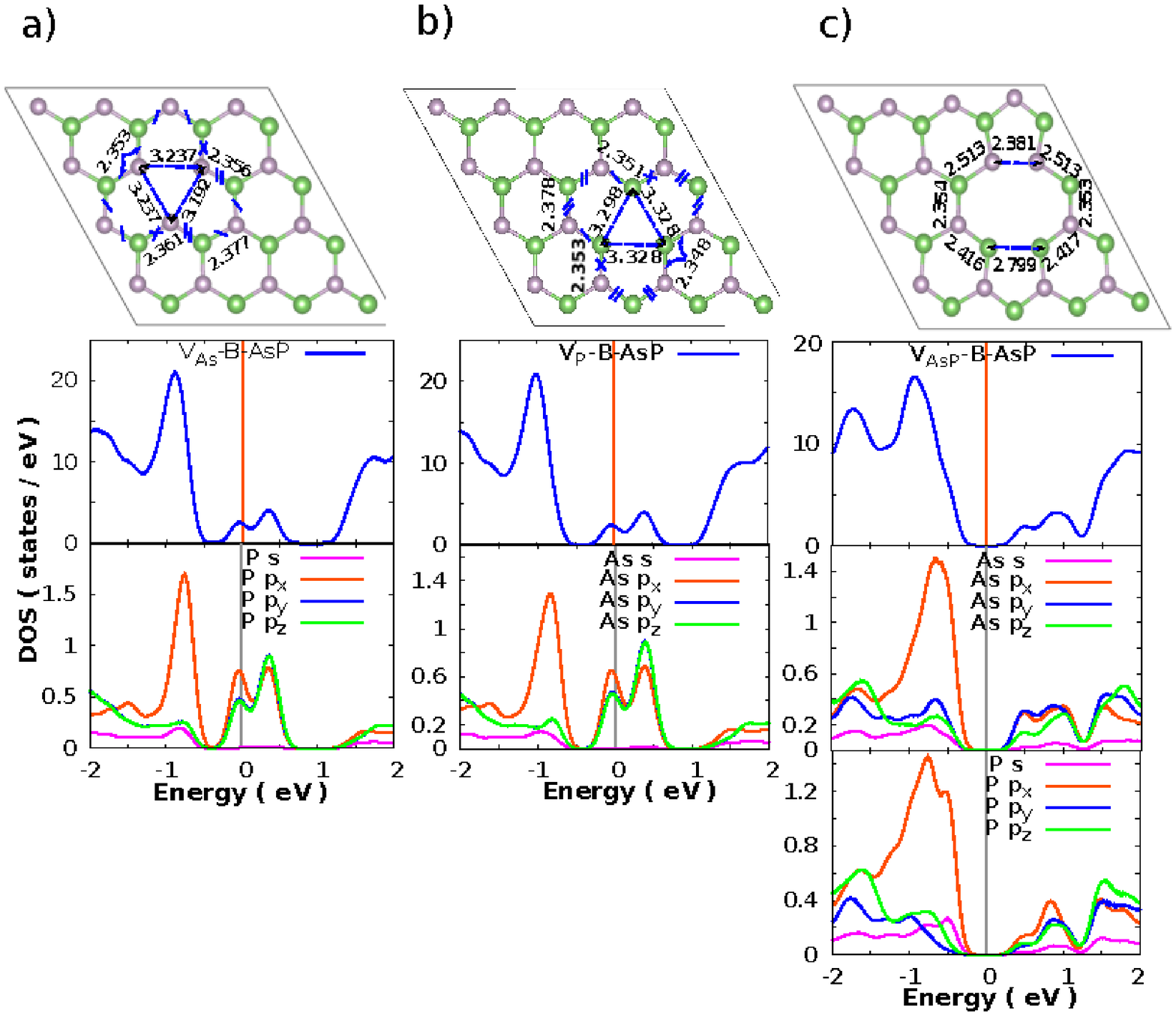}
\caption{Relaxed geometries for the 4x4 supercells,  the total and 
projected density of states plots for the defected structures on the B-AsP monolayer (a) $V_{As}$  b) $V_P $ c) $V_{AsP} $. 
In DOS plots, the total DOS of pristine B-AsP monolayer and the defected structure are illustrated together for 
better comparison. The following orbital
projected DOS plots correspond to atoms around the defect.  
All charge density plots are gerenerated by the VESTA \cite{VESTA}  program.  }
\label{fig:3}
\end{figure*}

Firstly, we examine the As vacany defect structure, which the fully relaxed 
structure is presented in Fig.\ref{fig:3}a. After optimization, 
three P atoms around the vacany are 
approaching  to each other, so  the P-P distance changed to 3.237 \AA\
 and to 3.192 \AA\
 which in the pristine state was 3.460 \AA. We found the bond length of As-P distance as
approximately 2.361 \AA\ which differs from the pristine monolayer while in pristine 
monolayer the As-P distance was 2.393 \AA. The accurate structural distances 
are shown in figures in detail. As shown in
Table\ref{tab:2}, since the cohesive energies of vacancies are almost equal, 
formation energy of $V_{As}$ is the highest among the others. To determine
the electronic properties of the system we obtained the density of states
(DOS) as presented in Fig.\ref{fig:3}a. As vacany induces a very small local
magnetization (0.04 $\mu_B$) and the system turns to metallic character.
This is the only defected state among the others on B-AsP monolayer
which turns into magnetic state. As seen in Fig.\ref{fig:3} in detail
on can concluded that two defect states appear in the band gap. While one
occurs from contributions of $p_x-p_y$ and $p_z$  orbitals of P atoms around the
vacany, the other state at the higher energy level are mixed state of
$s$, $p_x-p_y$ and $p_z$  orbitals of P atoms around vacany. 
According to L\"{o}wdin charge analysis \cite{Lowdin}, 
the excess charge on P atoms around vacany, which is calculated by substracting the charge at the atom from 
its valance charge, is 0.03 electrons while the other P atoms 
have 0.01  electrons excess charge in the bare state. 

The second defected structure under consideration is P vacany ($V_p$) which 
three dangling bonds occur with As-As = 3.328 \AA\ distance, as seen in 
Fig.\ref{fig:3}b. All the other bond lengths are shown in figure. 
This defected state is nonmagnetic and shows also metallic 
character. Like the $V_{As}$ case, two defect states occur at the band gap, 
while first consist of $p_x, p_y, p_z$ orbitals of As atoms around the 
vacany and the other has also contribution from the $s$ orbitals of the 
As atoms around vacany. According to L\"{o}wdin charge population analysis,
As atoms around vacany gives 0.06 electrons to the surrounding substrate. 
Additionallyi the most energetically favourable vacany defected state is the 
$V_P$ state.

Energetically second favourable vacany defected state among the considered
states is the divacany $V_{AsP}$, which is the state of lacking both
As and P atoms in the network. After the geometric optimization the hexagonal
structure is changed to an octagon and two pentagons, which is illustrated
in Fig.\ref{fig:3}c with detailed bond length informations. 
This is also the case where we have the biggest 
lattice correction with respect to the bare monolayer.
The $V_{AsP}$ system has 0.847 eV band gap due to the defect states. As shown
 in Fig.\ref{fig:3}c these defects levels  at around -0.75 eV and +0.70 eV 
  have significant contributions from $As_p$ and $P_p$ orbitals. 
L\"{o}wdin analysis
indicates that 0.10 electrons are transferred from As atoms 
 and 0.001 electrons are transferred to P atoms around the vacany. 

\begin{figure*}[t]
\hspace*{-2.50cm}\includegraphics[scale=0.85]{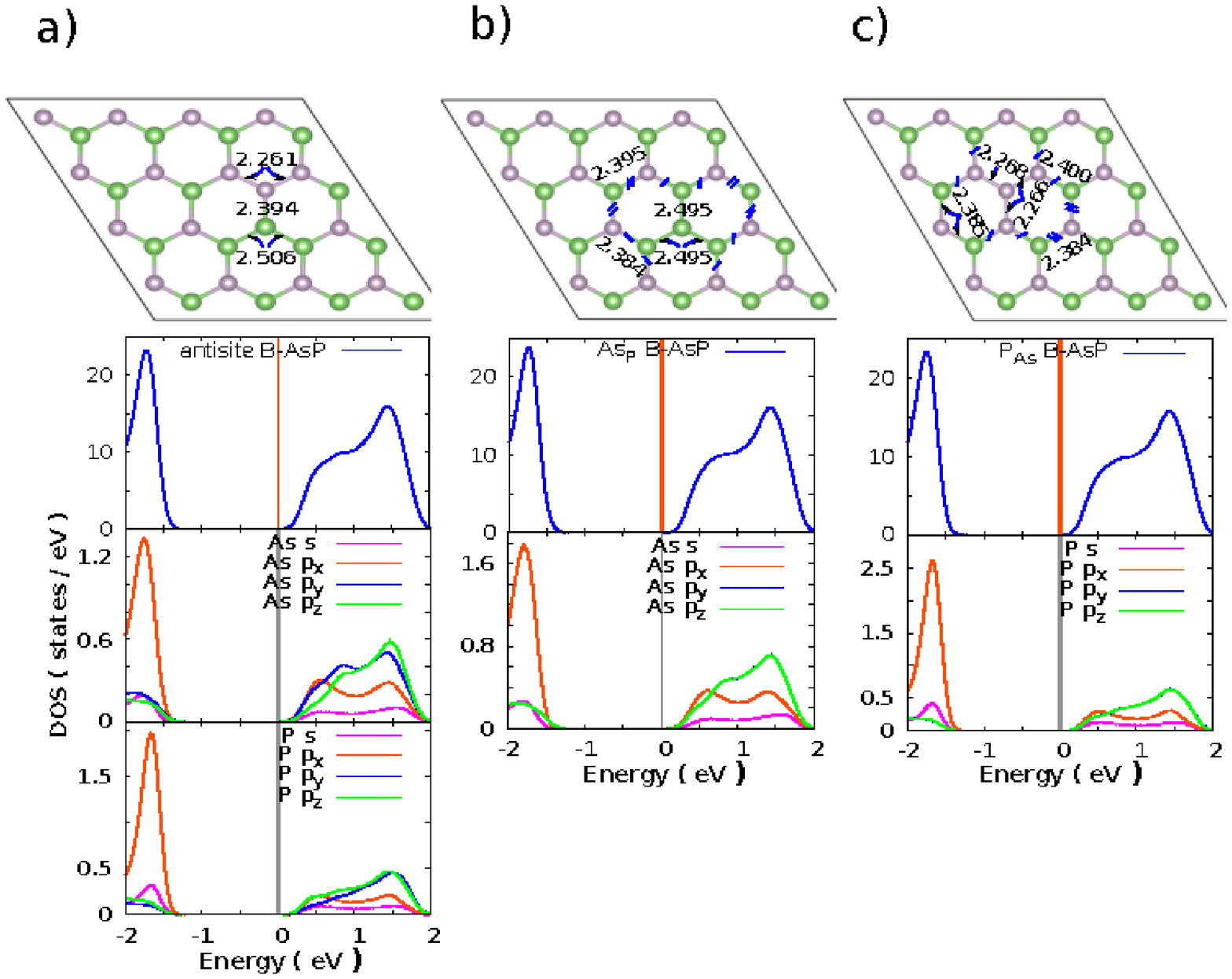}
\caption{  Relaxed geometries for the 4x4 supercells, the total and
projected density of states plots for the defected structures on the B-AsP monolayer
(a) $As \leftrightarrow P$  b) $As_P $ c) $P_{As} $.
In DOS plots, the total DOS of pristine B-AsP monolayer and the defected structure are illustrated together for
better comparison. The following orbital
projected DOS plots correspond to atoms around the defect.  }
\label{fig:4}
\end{figure*}

The other kind of system has antisite defect, which we exchange the positions 
of the neighbouring As and P atoms ($As \leftrightarrow P$) in the network
is presented in Fig.\ref{fig:4}a. We have detected that negligible bond length
elongation occur in that system. Also there is no lattice correction. The 
$d_{As-P}$ bond length changes only from 2.393 \AA\ to 2.394 \AA.  The P-P bond distance
and As-As bond distance are 2.261 \AA\ and 2.506 \AA\, respectively. Therefore 
the band gap is also affected minimal only slightly from 1.858 eV to 1.831 eV.
As the other cases the orbital contributions of As and P atoms around 
defects are shown in  Fig.\ref{fig:4}a. The excess charges on the exchanged
atoms are 0.11   and 0.02 electrons, respectively. These results with its small
formation energy implies that this structure can be occur or dominate defect 
during the fabrication of AsP monolayer.

The other kind of defected is obtained by substitution of As atom instead of
P atom in the monolayer ($As_P$). In the fully optimized structure of $As_P$
(Fig.\ref{fig:4}b, four As atoms are reorganized and they bonds to each other
with a bond length of As-As = 2.495 \AA. The As-P bond length is nearly  
the same with  the pristine system (2.395 \AA\ and 2.384 \AA\ 
which are denoted in the figure). 
There is no lattice correction. Also
this defect cannot induce any magnetization and the band gap changes  very 
slightly only to 1.804 eV. While the As atom at the central gives 0.08 
electrons to the surrounding the others give 0.12 electrons.

Last considered defect for B-AsP monolayer is $P_{As}$, the As atom 
is replaced by P atom in the network, is shown in Fig.\ref{fig:4}c with its
optimized structure.
This structure  has also similar geometry with the previous defect, where
the bond length of As-P is again 2.384 \AA\ and the P-P distance is approximately 2.268 \AA.
Also $P_{As}$ structure is nonmagnetic and is semiconductor with a band gap
1.842 eV. In Fig.\ref{fig:4}c, we illustrate the electronic DOS of 
$P_{As}$ which overlaps with  $As_P$ defect structure 
According to the L\"{o}wdin analysis, 
the central P atom has 0.16  electrons, while the other P atoms has -0.01 
electrons excess charges.

\subsection{Defect Structures on Pu-AsP monolayer}

\begin{figure*}[t]
\hspace*{-3.00cm}\includegraphics[scale=0.43]{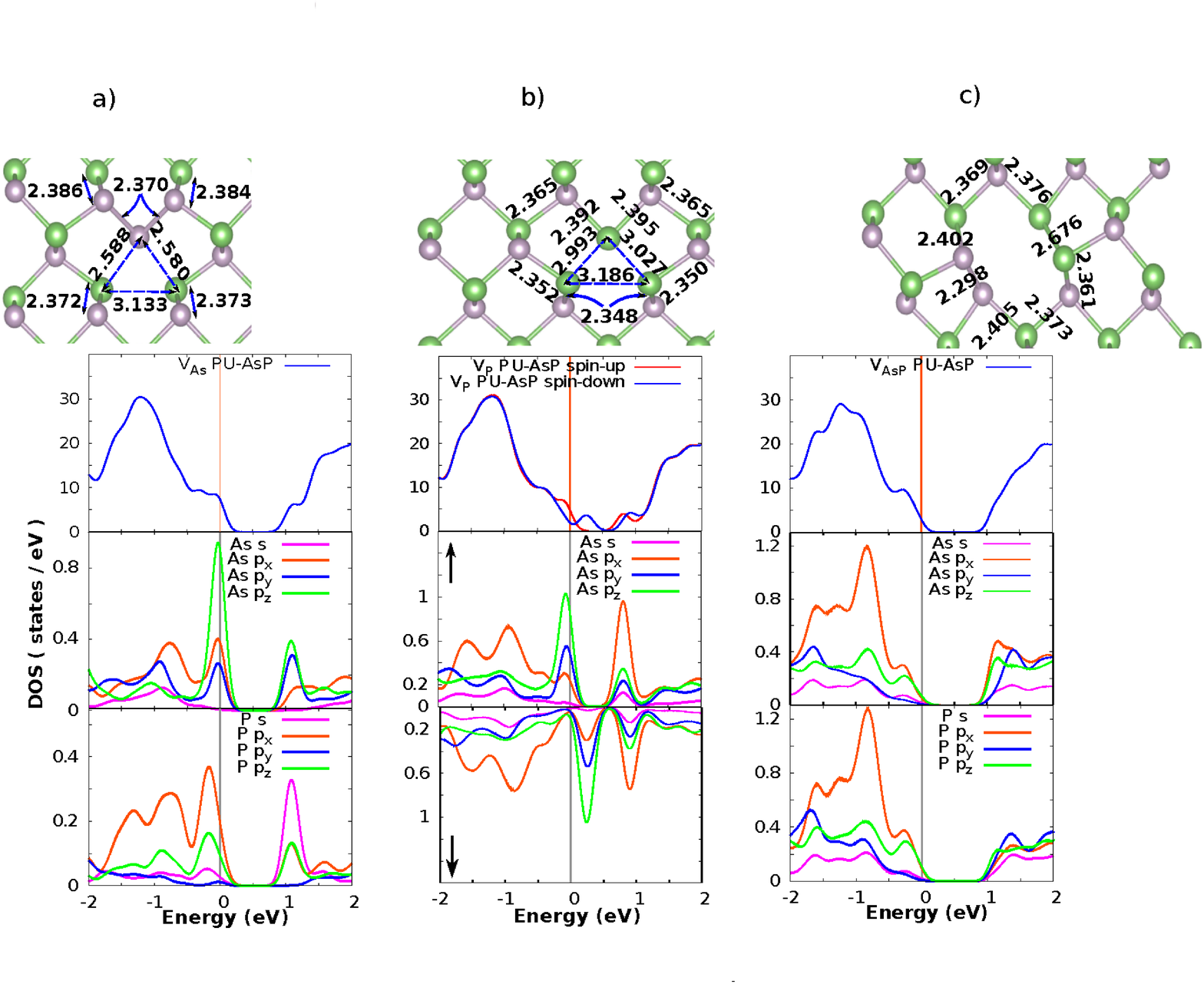}
\caption{Relaxed geometries for the 4x4 supercells, the total and
projected density of states plots for the defected structures on the Pu-AsP monolayer (a) $V_{As}$  b) $V_P $ c) $V_{AsP} $.
In DOS plots, the total DOS of pristine B-AsP monolayer and the defected structure are illustrated together for
better comparison. The following orbital
projected DOS plots correspond to atoms around the defect.   }
\label{fig:5}
\end{figure*}

As in the B-AsP monolayer case, the first considered vacany defect on Pu-AsP monolayer is
As vacany ($V_{As}$), illustrated in Fig.\ref{fig:5}a. In this defect structure the lattice 
constants $a$ and $b$ are changed from 14.00   \AA\ to 13.95 \AA\ and from 18.72 \AA\ to
18.65 \AA, respectively. The bond length of As-P atoms increases  from 2.395 \AA\ to 
2.580 \AA. Other structural distances are shown in figure. 
With these structural parameters this state is one of the most defected state among the others
which one can also concluded from the formation energy. To discuss the electronic properties
we obtained the density of states as presented in Fig.\ref{fig:5}a. 
This system is nonmagnetic.
The defect states 
occurs at the Fermi level and has contributions from both P and As atoms 
around the
vacany. According to the L\"{o}wdin charge population analysis, the excess charge on the
 P atoms around the vacany is -0.03  electrons. 

In the $V_P$ structure, we detect also lattice correction  where the lattice constant are 
$a = 13.97 $ \AA\ and $b= 18.68$ \AA. The accurate lengths related with the 
structure are given in the figure. This state is   magnetic  with a 
magnetization
of 1.19 $\mu_B$. The band gap  significanly decreases from 0.924 eV to zero eV
  in spin-up channel while a defect state from the spin-down channel is about
 +0.25 eV  from contributions of the $p_y$ orbital 
of As atoms around the vacany. The As atoms around the vacany 
gives 0.08  to the surrounding 
network where the other As atoms give 0.16 electrons.

The most energetically favourable state among the vacancies are the $V_{AsP}$ structure 
with a formation energy value of 0.926 eV ( Table\ref{tab:2}). After relaxation, 
we have found the new lattice parameters as  $a = 13.94$ \AA\ and
 $b = 18.64$  \AA, respectively. The As-P   
bond distance is approximately 2.37  \AA, where the newly constructed P-P and As-As bond distances are 2.298 \AA\
and 2.676 \AA. In distinction to the other vacany defects structures, divacany structure 
is nonmagnetic, with a larger band gap of 1.086 eV with respect to the bare Pu-AsP monolayer
 as well as  with respect to the other defected structures of Pu-AsP. The defect states above and
below the Fermi level has main contributions from $As_p$ and $P_p$ orbitals 
around the vacany.
Each P (As) atom of P-P (As-As) bond has excess charge of -0.03 (0.11) electrons.

\begin{figure*}[t]
\hspace*{-3.0cm}\includegraphics[scale=0.50]{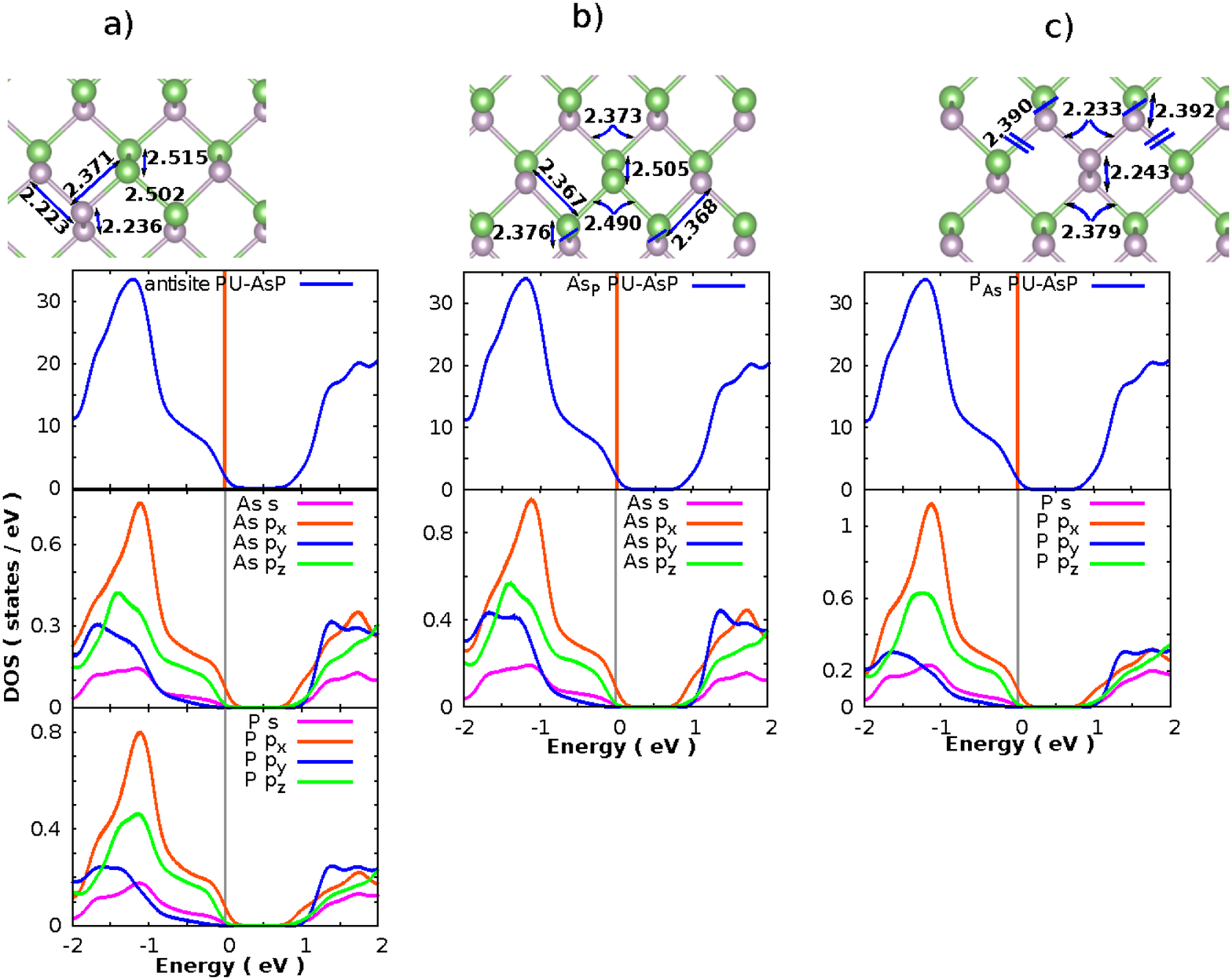}
\caption{ Relaxed geometries for the 4x4 supercells,  the total and
projected density of states plots for the defected structures on the Pu-AsP monolayer
(a) $As \leftrightarrow P$  b) $As_P $ c) $P_{As} $.
In DOS plots, the total DOS of pristine B-AsP monolayer and the defected structure are illustrated together for
better comparison. The following orbital
projected DOS plots correspond to atoms around the defect.  }
\label{fig:6}
\end{figure*}

The other kind of defect considered distinctively from the vacancies is the antisite defect where the P and As atoms
are exchanged (Fig.\ref{fig:6}a). We employ two different exchanging to the neighbouring P and As atoms. In the first type 
the exchanging is done on the x-direction and in the second one the exchanging is done on the y-direction. 
We detect no difference between these two defect structures  at  the end of the relaxation. On both system there are neither
lattice constant correction nor bond distance change between the P and As atoms. These defect structures are as stable as the
bare Pu-AsP monolayer structure is. The defect structures are nonmagnetic 
and have 0.91 eV band gaps. The exchanged atoms both give 0.06 electrons to the surrounding substrate. The formation
 energies are very small, so that there are no significant change due to these defect type. 

Next considered defect structures are $As_{P}$ and $P_{As}$ ( As (P) atom is placed instead of P (As) atom site) where 
both have minimal lattice corrections (Table\ref{tab:3}) have shown in Fig.\ref{fig:6}b and c with those related bond lengths. These structures are nonmagnetic
and have  0.912 eV and 0.921 eV band gaps, respectively. According to the L\"{o}wdin charge analysis, in $As_{P}$ ($P_{As}$) 
system the As (P) atom has excess charge of 0.03 (0.09)  electrons.
These defect structures have very similiar properties unlike their formation energies. From the formation energies $As_{P}$
defect is more dominate defect in comparison to $P_{As}$ defect.

\section{Conclusions}
In conclusion, based on the first principles plane wave calculation within the density functional theory, black and blue arsenic phosphorus monolayer structures are predicted.
 We show  the structural and electronic properties  of these monolayer structures. 
 The black AsP monolayer is found to be a direct semiconductor with a band gap of 0.924 eV
 while the blue counterpart has a indirect band gap of 1.804 eV. Additionally we   
investigated  the  electronic and magnetic properties of various kind of point defects including single and double
vacancies, antisite and substitutions on both puckered and buckled arsenic phosphorus. 
Our first principles density functional theory calculations show that the 
electronic and magnetic properties
of pristine B-AsP and Pu-AsP monolayers can be  adjustable by single and double vacancies, 
antisite and substition
defects. The nonmagnetic semiconductor nature of both monolayes can be transformed 
into metallic or narrow gap semiconductor upon
defected structures. The band gaps can be tuned
from metal character to 1.858 eV  for B-AsP and  
    from metal to 0.924 eV for Pu-AsP which
these ranges are especially important for radar and atmosferic applications.
Furthermore, the vacany defects induces magnetic moments  in both monolayers.

\section{Acknowledgement}
Computing resources used in this work were provided by the TUBITAK (The Scientific and Technical Research Council of Turkey) ULAKBIM, High Performance and Grid Computing Center (Tr-Grid e Infrastructure).

\newpage 
\textbf{Table 1:} Calculated parameters for  the two structures  corresponding to Pu-AsP  
and B-AsP   respectively. $a$, the lattice constant; $d(As-P)$, As-P distance; 
 $E$, total energy; $E_g$, band gap.
 
\textbf{Table 2:}  Calculated results for optimized defective structures of B-AsP
monolayer; $E_{coh}$, cohesive energy; $E_{form}$, formation energy, $a$,
 corrected lattice constant;
 $\mu$, magnetic moment per supercell in units of Bohr magneton $(\mu_B)$ (NM is for nonmagnetic systems).  
 
\textbf{Table 3:} Calculated results for optimized defective structures of Pu-AsP
 monolayer; $E_{coh}$, cohesive energy; $E_{form}$, formation energy, $a$ and
$b$, corrected lattice constants;
 $\mu$, magnetic moment per supercell in units of Bohr magneton $(\mu_B)$ (NM is for nonmagnetic systems).

\textbf{Fig.1:} Side view and top view of the structures of a) buckled B-AsP and b) puckered 
Pu-AsP.

\textbf{Fig.2:} Electronic band structures and corresponding density of states  the for structures  a) buckled B-AsP and b) puckered Pu-AsP. Zero of energy is set to Fermi level shown by dashed line. (For interpretation of the references to color in this figure legend, the reader is referred to the web version of this article.)

\textbf{Fig.3:} Relaxed geometries for the 4x4 supercells,  the total and 
projected density of states plots for the defected structures on the B-AsP monolayer (a) $V_{As}$  b) $V_P $ c) $V_{AsP} $. 
In DOS plots, the total DOS of pristine B-AsP monolayer and the defected structure are illustrated together for 
better comparison. The following orbital
projected DOS plots correspond to atoms around the defect. 

\textbf{Fig.4:} Relaxed geometries for the 4x4 supercells,  the total and 
projected density of states plots for the defected structures on the B-AsP monolayer 
(a) $As \leftrightarrow P$  b) $As_P $ c) $P_{As} $. 
In DOS plots, the total DOS of pristine B-AsP monolayer and the defected structure are illustrated together for 
better comparison. The following orbital
projected DOS plots correspond to atoms around the defect.

\textbf{Fig.5:} Relaxed geometries for the 4x4 supercells,  the total and 
projected density of states plots for the defected structures on the Pu-AsP monolayer (a) $V_{As}$  b) $V_P $ c) $V_{AsP} $. 
In DOS plots, the total DOS of pristine B-AsP monolayer and the defected structure are illustrated together for 
better comparison. The following orbital
projected DOS plots correspond to atoms around the defect.

\textbf{Fig.6:} Relaxed geometries for the 4x4 supercells,  the total and
projected density of states plots for the defected structures on the Pu-AsP monolayer 
(a) $As \leftrightarrow P$  b) $As_P $ c) $P_{As} $.
In DOS plots, the total DOS of pristine B-AsP monolayer and the defected structure are illustrated together for
better comparison. The following orbital
projected DOS plots correspond to atoms around the defect.

\newpage
\begin {table}\footnotesize
\caption{Calculated parameters for  the three structures  corresponding to Pu-AsP, B-AsPe and P-AsP respectively. 
$a$, the lattice constant; $d(As-P)$, As-P distance; }
\label{tab:1} 
\begin{center}
\begin{tabular}{lccccccccc} \\ \hline
\hline
 Structure  & &  & $a$ (\AA) & $b$ (\AA) & $d(As-P)$ (\AA)  & $E_{coh} $ (eV) & $E_{gap}$ (eV) &  &  \\ \hline 
\hline
$ Pu-AsP$ &   & & 3.50  & 4.68 &  2.382 & 4.782 & 0.924 &  &\\
 $B-AsP$ &   &  & 3.46  & - & 2.393 & 4.802 & 1.858 &  &  \\
\hline
\hline
\end{tabular} 
\end{center}
\end {table}

\newpage
\begin {table}\footnotesize
\caption{ Calculated results for optimized defective structures of B-AsP 
monolayer; $E_{coh}$, cohesive energy; $E_{form}$, formation energy, $a$,
 corrected lattice constant;
 $\mu$, magnetic moment per supercell in units of Bohr magneton $(\mu_B)$ (NM is for nonmagnetic systems).  }
\label{tab:2}
\begin{center}
\begin{tabular}{lccccccc} \\ \hline
\hline
  & $V_{As}$  & $V_{P}$ & $V_{AsP}$ & $As\leftrightarrow P$ & $As_{P}$ & $P_{As} $ \\ \hline 
\hline
$E_{coh} $(eV)  & 4.729 & 4.734 & 4.719 & 4.806 & 4.783 & 4.826 \\
$E_{form}$ (eV)  & 2.576 & 1.708 & 2.369 & 0.140 & -0.092 & -0.084 \\
$a$(\AA)   & 13.70 & 13.71 & 13.58 & 13.84 & 13.84 & 13.78 \\
 $\mu (\mu_B)$   & 0.04 & NM & NM & NM & NM & NM\tabularnewline
\hline
\hline
\end{tabular}
\end{center}
\end {table}

\newpage
\begin {table}\footnotesize
\caption{ Calculated results for optimized defective structures of Pu-AsP
 monolayer; $E_{coh}$, cohesive energy; $E_{form}$, formation energy, $a$ and 
$b$, corrected lattice constants;
 $\mu$, magnetic moment per supercell in units of Bohr magneton $(\mu_B)$ (NM is for nonmagnetic systems).  }
\label{tab:3}
\begin{center}
\begin{tabular}{lccccccc} \\ \hline 
\hline
  & $V_{As}$  & $V_{P}$ & $V_{AsP}$ & $As\leftrightarrow P$ & $As_{P}$  & $P_{As}$ \\ \hline 
\hline
$E_{coh}$ (eV) & 4.759 & 4.754 &  4.765   &  4.784 & 4.772 & 4.795 \\
$E_{form}$ (eV)  &  1.739 & 1.336  &  0.926 & -0.115  &  -0.052 &  1.336 \\ 
$a$ (\AA)  & 13.95 & 13.97 &  13.94   &  14.00 & 14.02 & 14.00 \\
$b$ (\AA) & 18.65 & 18.68 &  18.64   &  18.72 & 18.74 & 18.72 \\
 $\mu (\mu_B)$  & NM & 1.19  &  NM &  NM &  NM & NM \tabularnewline
\hline
\hline
\end{tabular}
\end{center}
\end {table}

\newpage
\pagebreak

\newpage
\begin{thebibliography}{00}

\bibitem{Geim}
 Geim, A. K.; Novoselov, K. S. The Rise of Graphene. Nat. Mater. 2007, 6, 183-191.


\bibitem{Novoselov}
 Novoselov, K. S.; Geim, A. K.; Morozov, S. V.; Jiang, D.; Zhang, Y.; Dubonos, S. V.; Grigorieva, I. V.; Firsov, A. A. Science 2004, 306, 666-669.

\bibitem{Durgun}
 E. Durgun, S. Tongay, and S. Ciraci, Phys. Rev. B 72, 075420 (2005).

\bibitem{Cahangirov}
 S. Cahangirov, M. Topsakal, E. Aktürk, H. Sahin, and S. Ciraci, Phys. Rev. Lett. 102, 236804 (2009).

\bibitem{Vogt} 
P. Vogt, P. Depadova, C. Quaresima, J. Avila, E. Frantzeskakis, M. C. Asensio, A. Resta, B. Ealet, and G. Lelay, Phys. Rev. Lett. 108, 155501 (2012).

\bibitem{Ozcelik}
 V. O. Ozçelik, E. Durgun, and S. Ciraci, J. Phys. Chem. Lett. 5, 2694 (2014).

\bibitem{Xu} 
Y. Xu, B. Yan, H. J. Zhang, J. Wang, G. Xu, P. Tang, W. Duan, and S. C. Zhang, Phys. Rev. Lett. 111, 136804 (2013).

\bibitem{Sahin}
 H. Sahin, S. Cahangirov, M. Topsakal, E. Bekaroglu, E. Akturk, R. T. Senger, and S. Ciraci, Phys. Rev. B 80, 155-453 (2009).

\bibitem{Tongay} 
S.Tongay, S. Dag, E. Durgun, R. T. Senger, and S. Ciraci, J. Phys.: Condens. Matter 17, 3823 (2005).

\bibitem{Malko}
 D. Malko, C. Neiss, F. Vines, and A. Görling, Phys. Rev. Lett. 108, 086804 (2012).

\bibitem{Ozcelik2} 
V. O. Ozçelik and S. Ciraci, J. Phys. Chem. C 117, 2175 (2013).

\bibitem{Ozcelik3} 
V. O. Ozçelik, S. Cahangirov, and S. Ciraci, Phys. Rev. Lett. 112, 246803 (2014).

\bibitem{Xu2} 
X. Xu, J. Zhuang, Y. Du, H. Feng, N. Zhang, C. Liu, T. Lei, J. Wang, M. Spencer, T. Morishita, X.  Wang, and S. X. Dou, Sci. Rep. 4, 7543 (2014).

\bibitem{Joensen} 
P. Joensen, R. F. Frindt, and S. R. Morrison, Mater. Res. Bull. 21, 457 (1968).

\bibitem{Coleman} 
J. N. Coleman, M. Lotya, A. O. Neil, S. D. Bergin, P. J. King, U. Khan, and K. Young, Science 331, 568 (2011).

\bibitem{Ataca1} 
C. Ataca and S. Ciraci, J. Phys. Chem. C 115, 13303 (2011).

\bibitem{Ataca2} 
C. Ataca, H. Sahin, and S. Ciraci, J. Phys. Chem. C 116, 8983 (2012).


\bibitem{Bekaroglu}
E. Bekaroglu, M. Topsakal, S. Cahangirov, S. Ciraci,  Phys. Rev. B 81 (2010) 075433.


\bibitem{Sun}
L. Sun, Y. Li, Z. Li, Q. Li, Z. Zhou, Z. Chen, J. Yang, J.G. Hou, J. Chem. Phys. 129 (2008) 174114.

\bibitem{Xu3}
B. Xu, J. Yin, Y.D. Xia, X.G. Wan, Z.G. Liu,  Appl. Phys. Lett. 96 (2010) 143111.



\bibitem{Li} L. Li, Y. Yu, G. J. Ye, Q. Ge, X. Ou, H. Wu, D. Feng, X. H. Chen, and Y. Zhang, Nat. Nanotechnol. 9, 372-377 (2014).

\bibitem{Zhu} 
Z. Zhu and D. Tomanek, Phys. Rev. Lett. 112, 176802 (2014).

\bibitem{Low} 
T. Low, A. S. Rodin, A. Carvalho, Y. Jiang, H. Wang, F. Xia, and A. H. Castro Neto, Phys. Rev. B 90, 075434 (2014).



\bibitem{PRL}
Z. Zhu  and D. Tomanek, Phys. Rev. Let. 112 (2014) 176802.


\bibitem{PRBarsenene}
C. Kamal  and M. Ezawa,  Phys. Rev. B 91 (2015) 085423.


\bibitem{PRBakturk}
O. \"{U}zengi Akturk, V. Ongun \"{O}z\c{c}elik, S. Ciraci, Phys. Rev. B 91 (2015) 234446.


\bibitem{AngeChemie}
S. Zhang, Z. Yan, Y. Li, Z. Chen, H. Zeng, Angew. Chem. Int. Ed. 54 (2015) 3112, 3115.

\bibitem{Zhang}
 Y. Zhang , T. T. Tang , C. Girit , Z. Hao , M. C. Martin , A. Zettl , M. F. Crommie , Y. R. Shen , F. Wang , Nature 2009 , 459 , 820 .

\bibitem{Narita}
 Narita, S.; Akahama, Y.; Tsukiyama, Y.; Muro, K.; Mori, S.; Endo, S.; Taniguchi, M.; Seki, M.; Suga, S.; Mikuni, A.; Kanzaki, H. Physica B+C 1983, 117-118, 422-424.

\bibitem{Maruyama} 
Maruyama, Y.; Suzuki, S.; Kobayashi, K.; Tanuma, S. Physica B+C 1981, 105, 99-102.

\bibitem{Liu} 
Liu, H.; Neal, A. T.; Zhu, Z.; Luo, Z.; Xu, X.; Tomanek, D.; Ye, P. D. ACS Nano 2014, 8, 4033-4041.

\bibitem{Zhu2} 
Zhu, Z.; Guan, J.; Tomanek, D. Phys. Rev. B: Condens. Matter Mater. Phys. 2015, 91, 161404.

\bibitem{Liu2}
 Liu, B.; Köpf, M.; Abbas, A. N.; Wang, X.; Guo, Q.; Jia, Y.; Xia, F.; Weihrich, R.; Bachhuber, F.; Pielnhofer, F.; Wang, H.; Dhall, R.; Cronin, S.; Ge, M.; Fang, X.; Nilges, T.; Zhou, C. Adv. Mater. 2015, 27, 4423-4429.

\bibitem{Zhang2} 
Zhang, Y.; Tan, Y. W.; Stormer, H. L.; Kim, P. Nature 2005, 438, 201-204.

\bibitem{Koenig} 
S. P. Koenig, R. A. Doganov, H. Schmidt, A. H. Castro Neto and B. özyilmaz, Appl. Phys. Lett., 2014, 104, 103106.

\bibitem{Gomez} 
A.Castellanos-Gomez, L. Vicarelli, E. Prada, J. O. Island, K. L. Narasimha-Acharya, S. I. Blanter, D. J. Groenendijk, M. Buscema, G. A. Steele, J. V. Alvarez, H. W. Zandbergen, J. J. Palacios and H. S. J. van der Zant, 2D Mater., 2014, 1, 025001.

\bibitem{Xia} 
F. Xia, H. Wang and Y. Jia, Nat. Commun. 2014, 5, 4458.

\bibitem{Wang} 
X. M. Wang , A. M. Jones , K. L. Seyler , V. Tran , Y. C. Jia , H. Zhao , H. Wang , L. Yang , X. D. Xu , F. N. Xia , Nat. Nanotechnol. 2015 , DOI:10.1038/nnano.2015.71.

\bibitem{Tomanek} 
Z. Zhu, J. Guan and D. Tomanek, Nano Lett., 2015, 15, 6042.


\bibitem{Adv.Mat.2015}
B. Liu, M. K\u{o}pf, A. Abbas, X. Wang, Q. Guo, Y. Jia, F. Xia, R. Weihrich, F. Bachhuber, F. Pielnhofer, H. Wang, R. Dhall, S. Cronin, M. Ge, X. Fang, T. Nilges, C. Zhou,  Adv. Mater.  27 (2015) 4423-4429.



\bibitem{paw}
P.E. Bl\"ochl,  Phys. Rev. B 50 (1994) 17953.

\bibitem{pbe}
J.P. Perdew, K. Burke, M. Ernzerhof,  Phys. Rev. Lett. 77 (1996) 3865.

\bibitem{pwscf}
P. Giannozzi, S. Baroni, N. Bonini, M. Calandra, R. Car, C. Cavazzoni, D. Ceresoli, G.L. Chiarotti, M. Cococcioni, I. Dabo, A. Dal Corso, S. Fabris, G. Fratesi, S. de Gironcoli, R. Gebauer, U. Gerstmann, C. Gougoussis, A. Kokalj, M. Lazzeri, L. Martin-Samos, N. Marzari, F.  Mauri, R. Mazzarello, S. Paolini, A. Pasquarello, L. Paulatto, C. Sbraccia, S. Scandolo, G. Sclauzero, A.P. Seitsonen, A. Smogunov, P. Umari, R.M. Wentzcovitch, QUANTUM ESPRESSO: A Modular and Open-Source Software Project for Quantum Simulations of Materials, J. Phys: Condens. Matter 21 (2009) 395502.

\bibitem{monk}
H.J. Monkhorst, J.D. Pack,  Phys. Rev. B 13 (1976) 5188.

\bibitem{BFGS}
C.G. Broyden, The Convergence of a Class of Double-rank Minimization Algorithms 1. General Considerations, Journal of Institute of Mathematics and Its Applications 6 (1970) 76-90.


\bibitem{Lowdin}
P.-O. L\"{o}wdin, J. Chem. Phys. 18 (1950) 365.



\bibitem{VESTA}
K. Momma, F. Izumi, VESTA 3 for three-dimensional visualization of crystal, volumetric and morphology data, J. Appl. Crystallogr. 44 (2011) 1272–1276.



\end{thebibliography}
\end{document}